\documentclass[aps,prd,reprint,preprintnumbers,superscriptaddress,showpacs,twocolumn]{revtex4}
\usepackage{latexsym,graphicx,amssymb,amsmath,mathrsfs}
\usepackage{setspace,bm}
\usepackage[colorlinks=true, pdfstartview=FitV, linkcolor=red, citecolor=blue, urlcolor=blue]{hyperref}
\usepackage[caption=false]{subfig}

\usepackage{amsmath}
\usepackage{ulem}
\usepackage[usenames]{color}

\begin{document}
\preprint{YITP-15-63, KUNS-2571}

\title{Inhomogeneous chiral condensates and non-analyticity\\
       under an external magnetic field}

\author{Kouji Kashiwa}
\email[]{kouji.kashiwa@yukawa.kyoto-u.ac.jp}
\affiliation{Yukawa Institute for Theoretical Physics,
Kyoto University, Kyoto 606-8502, Japan}

\author{Tong-Gyu Lee}
\email[]{lee@ruby.scphys.kyoto-u.ac.jp}
\affiliation{Department of Physics, Kyoto University,
Kyoto 606-8502, Japan}

\author{Kazuya Nishiyama}
\email[]{nishiyama@ruby.scphys.kyoto-u.ac.jp}
\affiliation{Department of Physics, Kyoto University,
Kyoto 606-8502, Japan}

\author{Ryo Yoshiike}
\email[]{yoshiike@ruby.scphys.kyoto-u.ac.jp}
\affiliation{Department of Physics, Kyoto University,
Kyoto 606-8502, Japan}

\begin{abstract}
We investigate inhomogeneous chiral condensates,
 such as the so-called dual chiral density wave of dense quark matter,
 under an external magnetic field
 at finite real and imaginary chemical potentials.
In a model-independent manner,
 we find that analytic continuation
 from imaginary to real chemical potential is not possible
 due to the singularity induced by inhomogeneous chiral condensates at
 zero chemical potential.
From the discussion on the non-analyticity
 and methods used in lattice QCD simulations,
 e.g., Taylor expansion,
 and the analytic continuation with an imaginary chemical potential,
 it turns out that information on an inhomogeneous chiral condensed phase is missed
 in the lattice simulations at finite baryon chemical potentials
 unless the non-analyticity at zero chemical potential is correctly
 considered.
We also discuss an exceptional case without such non-analyticity
 at zero chemical potential.
\end{abstract}

\maketitle

\section{Introduction}
Exploring the phase diagram of Quantum Chromodynamics (QCD)
 is one of the important and interesting subjects
 in nuclear physics, elementary particle physics, and astrophysics;
for a review on the QCD phase diagram see Ref.~\cite{Fukushima:2010bq}.
{\it Ab initio} lattice QCD simulations are a powerful and
 reliable method for analyzing the nonperturbative nature of QCD,
 but the well-known sign problem arises at finite real chemical potential
 ($\mu_\mathrm{R}$).
To circumvent the sign problem, several methods have been proposed so far:
 the Taylor expansion method, the reweighting method, the canonical approach,
 and the analytic continuation method
 from imaginary chemical potentials ($\mu_\mathrm{I}$); see for example
 Ref.~\cite{deForcrand:2010ys}.
These methods are, however, limited in the $\mu_\mathrm{R}/T < 1$ region,
 where $T$ is the temperature.
On the other hand, effective model calculations are useful
 to investigate the phase diagram at finite $T$ and $\mu_\mathrm{R}$,
 but it is still difficult to quantitatively discuss the phase structure of QCD
 due to a large ambiguity of the model.

Inhomogeneous chiral condensed phases have been vigorously studied
 in the QCD phase diagram at finite density
 within chiral effective models,
 and also supported from Dyson-Schwinger studies of dense QCD
 (for a recent review see Ref.~\cite{Buballa:2014tba}),
 and their stabilities against thermal fluctuations
 have also been discussed~\cite{Lee:2015bva,Hidaka:2015xza}.
These phases have the quark condensate
 taking a spatially inhomogeneous configuration.
One example with a one-dimensional modulation includes
 the so-called dual chiral density wave (DCDW)~\cite{Nakano:2004cd},
 where the chiral order parameter is spatially modulated
 with a finite wavenumber ($q$) in the single direction.
Several studies based on mean-field calculations predict
 that the inhomogeneous phase can be realized
 in sufficiently high $\mu_\mathrm{R}$ region.
However, this situation could be changed
 if we introduce an external magnetic field.

Magnetic aspects of QCD have been attracted much attention
 in the physics of heavy-ion collision experiments
 and compact stars (for a review see Ref.~\cite{Kharzeev:2012ph}).
Recently, it has been reported that quark matter with the DCDW condensate
 could exhibit spontaneous magnetization~\cite{Yoshiike:2015tha},
 which may be related to the origin of strong magnetic fields in compact
 stars.
In the presence of the external magnetic field,
 the most important and interesting point is that
 the DCDW phase appears even at small values of $\mu_\mathrm{R}$
 except $\mu_\mathrm{R}=0$
 within the massless Nambu--Jona-Lasinio (NJL)
 model~\cite{Frolov:2010wn,Tatsumi:2014wka,Nishiyama:2015fba}.
This naturally raises the following question:
 Can we investigate inhomogeneous chiral condensates
 by using dense lattice QCD simulations at finite $\mu_R$
 when the DCDW phase appears at very small $\mu_\mathrm{R}$?
Here we assume that the inhomogeneous phase can appear
 in finite and discretized systems.

In this paper, we discuss spatially inhomogeneous chiral condensates
 and the associated non-analyticity.
Then we show the possibility that information on
 the inhomogeneous phases can be missed
 in the lattice QCD simulations
 with the Taylor expansion method and the analytic continuation method
 from the $\mu_\mathrm{I}$ region to the $\mu_\mathrm{R}$ region.
We call this problem the ``information missing problem.''
We find that
 the applicable range of $\mu_\mathrm{R}$
 in the dense lattice QCD simulations at finite $\mu_R$
 is strongly restricted, and such numerical simulations
 cannot describe the correct system
 with inhomogeneous chiral condenses
 if we neglect the non-analytic properties at $\mu^2=0$.
Some problems in the canonical and reweighting methods
 are also discussed.
In addition, an exceptional case
 where there is no non-analyticity at $\mu^2=0$ is presented.
Here note that the same problem would exist in zero magnetic field,
 because the DCDW phase can appear also
 in the case without the external magnetic field.

\section{Inhomogeneous chiral condensates at imaginary chemical potential}
In realistic systems,
 the grand canonical partition function ($Z$) must be real.
It is also valid in the presence of inhomogeneous chiral condensates
 under a constant magnetic field,
 since the ground states of inhomogeneous phases correspond to
 solutions of Hamiltonian with $\mu_\mathrm{R}$, i.e.,
 $Z(T,\mu_\mathrm{R},B;q) \in \mathbb{R}$,
 where $B$ is the strength of the external magnetic field.
This relation is simply manifested in the NJL model.
The same discussion in the Polyakov-loop-extended NJL model
 requires some extensions of the formalism,
 e.g., considering a complex path
 contour~\cite{Nishimura:2014rxa,Nishimura:2014kla,Tanizaki:2015pua}
 due to the model sign problem.

Here we only focus on the DCDW phase,
 since it is well known that the so-called real kink crystal (RKC) phase
 does not appear in the $\mu_\mathrm{I}$ region
 at least in the Gross-Neveu model~\cite{Karbstein:2006er}.
Also, the RKC phase shows no significant dependence on the
 external magnetic
 field and thus appears at sufficiently high $\mu_\mathrm{R}$,
 while the DCDW phase is extended to small $\mu_\mathrm{R}$ region.
Therefore, we do not here treat the RKC condensate,
 since we are interested in the inhomogeneous phase at small
 $\mu_\mathrm{R}$.

Assuming that the DCDW type inhomogeneous condensation arises
 in the presence of the uniform magnetic field,
 the massless NJL Lagrangian density with the mean-field
 approximation takes the form~\cite{Frolov:2010wn}
\begin{align}
 {\cal L} =
 \bar{ \psi} \left(i\gamma^\mu D_\mu +\mu\gamma^0 -m
 + \gamma^5 \gamma^3 \frac{q}{2} \right)\psi - \frac{m^2}{4G},
\end{align}
 where $\psi$ is the quark field,
 $\gamma^\mu$ are Dirac matrices,
 $\mu$ is the quark chemical potential,
 $G$ is the coupling constant,
 the constituent quark mass $m=-2G\Delta$
 with $\Delta$ the constant amplitude of the DCDW,
 and the covariant derivative $D_{\mu} = \partial_{\mu} - ieA_{\mu}$
 with $e$ the electric charge and $A_{\mu}$ the external magnetic field.
Here $A_{\mu}$ is directed along the $z$ axis, $\nabla\times{\bf A} = B\hat{\bf z}$.
The Lagrangian density given above is the one-flavor case
 but can be easily extended to the two-flavor one by considering the
 $\pi^0$-modulation.
Then the quark determinant in the partition function takes the form
\begin{align}
 \ln Z
 \sim {\rm Tr ln}
 \left(
 i\gamma^{\mu}D_{\mu} + \mu\gamma^0 - m + \gamma^5\gamma^3 q/2
 \right). \label{det}
\end{align}
Here we can show the symmetry of the system model-independently,
 while the analysis described above is model dependent.
To this end, we evaluate $Z$ by a Ginzburg-Landau (GL) expansion.

In general, $q$-odd terms are not banned
 in the expansion of $Z$ with respect to $q$ under the constant magnetic field:
\begin{align}
 Z =
 \alpha_0 + \alpha_1 q + \alpha_2 q^2 + \alpha_3 q^3 + \alpha_4 q^4 + \cdots, \label{gl}
\end{align}
where $\alpha_n$ ($n=0,1,2,...$) are expansion coefficients
 determined by effective chiral models like the NJL model.
Here the $q$-odd terms can be induced by
 a scalar quantity (${\bf q}\cdot{\bf B}$)
 without violating the rotational invariance.
The coefficients $\alpha_{2n}$ ($\alpha_{2n+1}$)
 should be an even (odd) function of $B$ and $\mu$, respectively,
 which can be understood by taking the trace of Dirac space in Eq.~(\ref{det}).
This implies that $q$ of the DCDW can take an imaginary value
 in the finite $\mu_\mathrm{I}$ region so as to manifest the reality of
 $Z$, i.e., $Z(T,\mu_\mathrm{I},B;q=iq_{\rm I}) \in \mathbb{R}$
 where $q_{\rm I} \in \mathbb{R}$.
This situation can arise in the case where
 an anomalous quark number density is proportional to
 $q$~\cite{Tatsumi:2014wka},
 since the quark number density ($n_q$) should be pure imaginary
 in the $\mu_\mathrm{I}$ region
 and $q$ acts as $n_q$.
A similar situation can also be expected,
 e.g., in complex chemical potential system,
 where $Z$ should be a complex value.

\begin{figure}[t]
\begin{center}
 \includegraphics[width=0.35\textwidth]{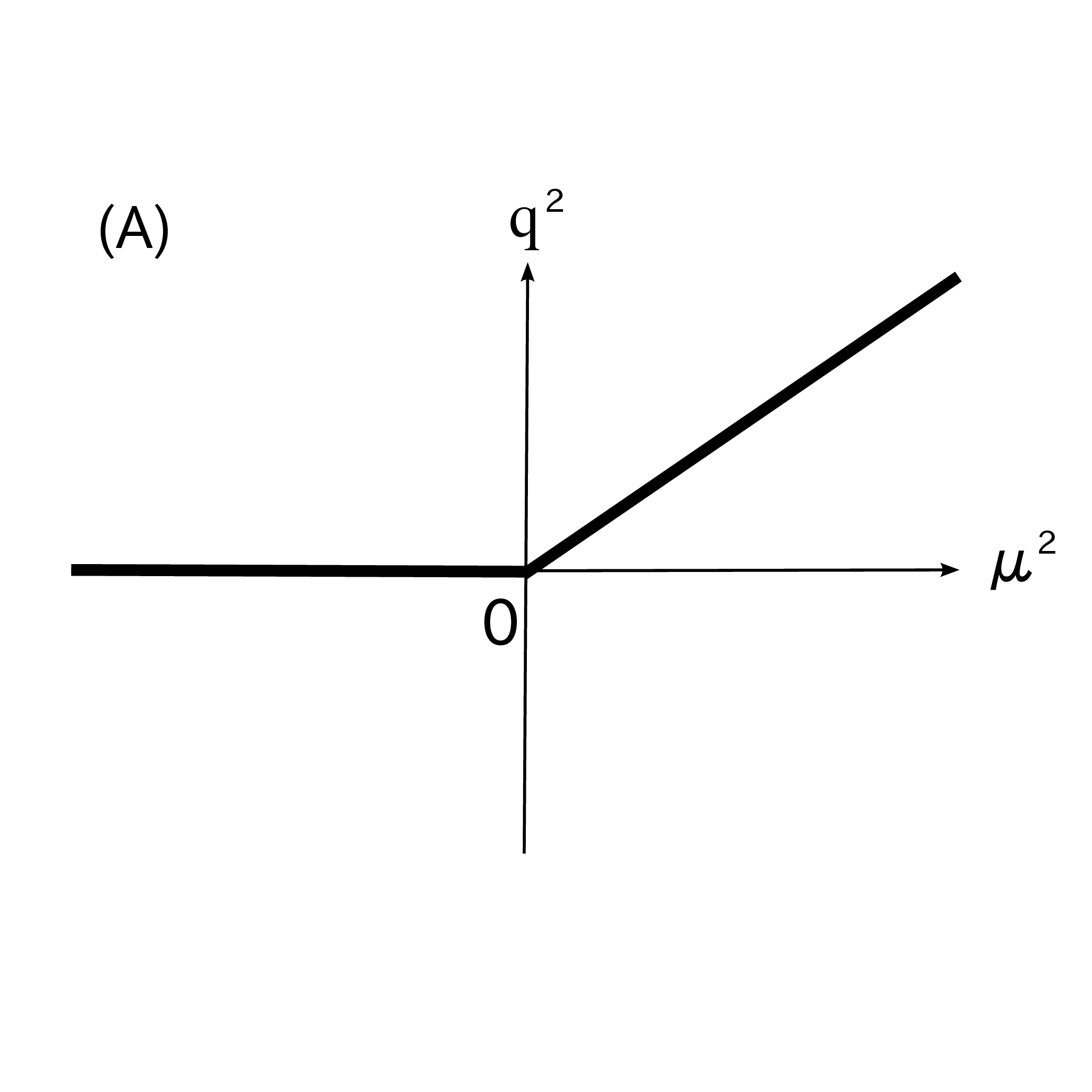}
 \includegraphics[width=0.35\textwidth]{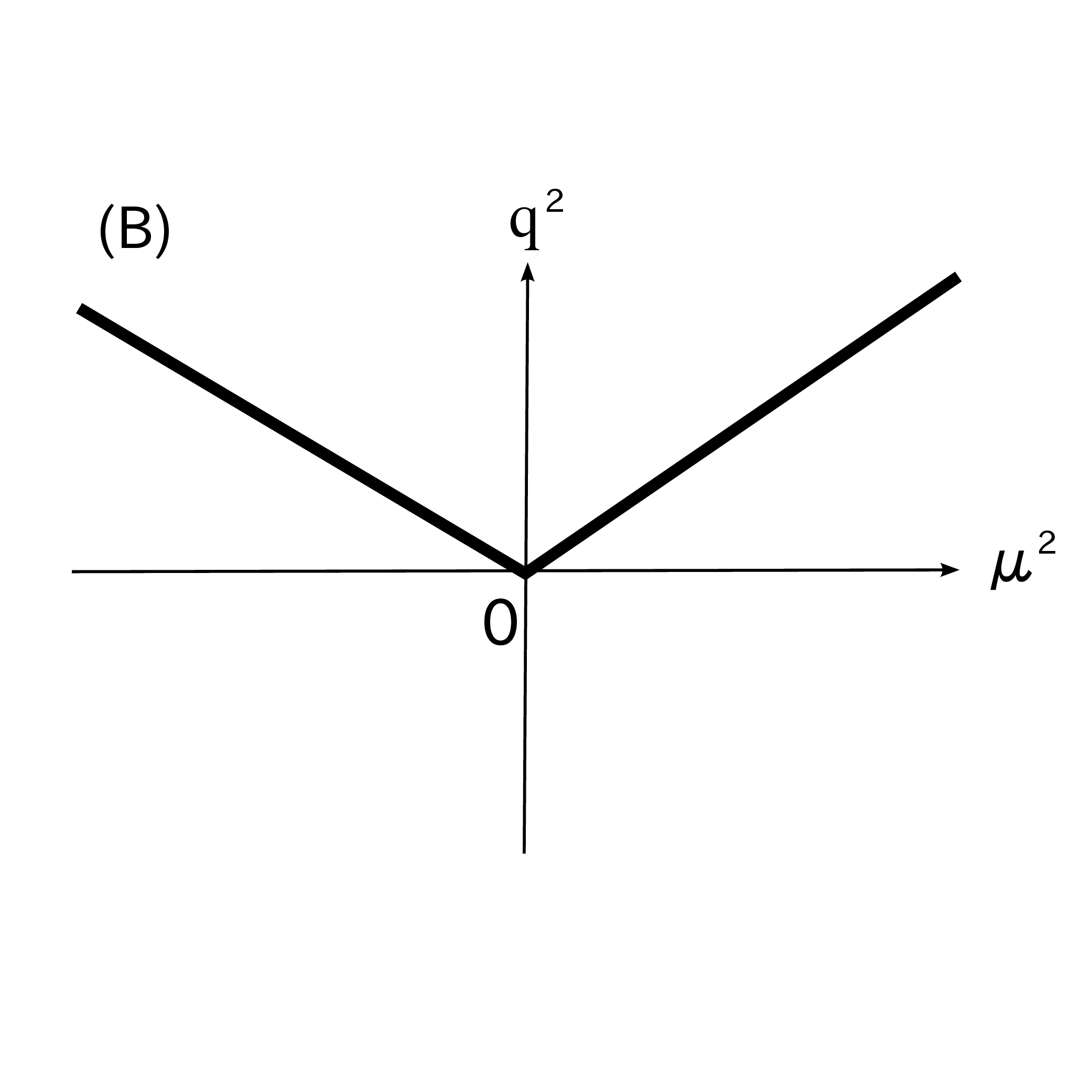}
 \includegraphics[width=0.35\textwidth]{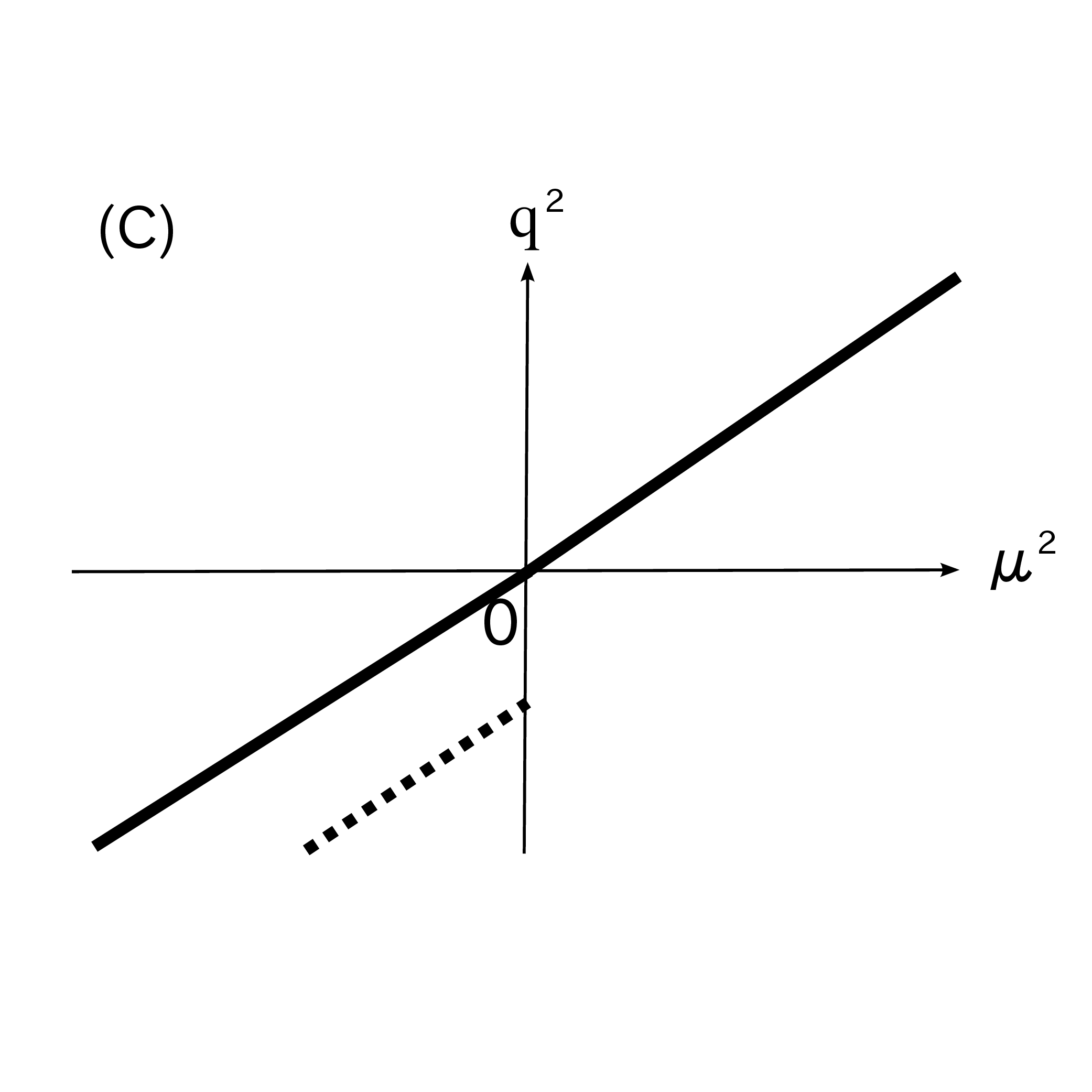}
\end{center}
\caption{
Schematic figures with respect to $q^2$ as a function of $\mu^2$
 at very close to $\mu^2=0$
 for each scenario.
The $\mu^2=0$ point becomes singular,
 except the solid lined case in Scenario C.
}
\label{Fig:Scenarios_A_B_C}
\end{figure}

By summarizing the above discussions, we can reach the following
scenarios:
\begin{description}
\item[Scenario A --- ]
 The DCDW phase does not appear at finite $\mu_\mathrm{I}$.
\item[Scenario B --- ]
 The DCDW phase appears at finite $\mu_\mathrm{I}$,
 but the partition function becomes complex.
\item[Scenario C ---]
 The DCDW phase appears at finite $\mu_\mathrm{I}$,
 and the partition function is real.
\end{description}
Each scenario is illustrated in Fig.~\ref{Fig:Scenarios_A_B_C}.
At a point very close to $\mu^2=0$,
 it is enough to regard $Z$ as
 only the first few terms in Eq.~(\ref{gl}),
 since we can consider that
 $q$ becomes quite small if there is no first-order transition at
 $\mu^2=0$.
In the region of $\mu^2>0$,
 the stationary condition for $q$, $\partial Z / \partial q = 0$,
 yields $q=-\alpha_1 / 2\alpha_2 \sim \mu$.
Therefore, $q^2$ is described as a linear function of $\mu^2$,
 as shown in Fig.~\ref{Fig:Scenarios_A_B_C}.
Here one can model-independently consider that $Z$ contains $\alpha_1$,
 since it is known in Ref.~\cite{Tatsumi:2014wka}
 that the GL coefficient $\alpha_1$ evaluated within the NJL model
 contains the Wess-Zumino-Witten (WZW) term~\cite{Son:2007ny}
 derived from chiral anomaly.
In Scenario B, the wave number is real at finite $\mu_\mathrm{I}$,
 $q \in \mathbb{R}$, while in Scenario C it becomes pure imaginary,
 $iq \in \mathbb{R}$.
Although we need a model calculation
 to see which of scenarios is realized,
 our purpose in this paper is to qualitatively understand
 inhomogeneous chiral condensates and their associated non-analyticities,
 so that we leave it for future work.
In the following, we discuss details of each scenario without model
 assumptions.

\subsection{Non-analyticity and analytic continuation}

We discuss the analytic continuation from the $\mu_\mathrm{I}$
 to the $\mu_\mathrm{R}$ region on the $\mu^2$ axis.
Here we introduce two holomorphic functions, $F_1$ and $F_2$,
 which describe regions
 $C_1 = \{ \mu^2 \in \mathbb{R} ~|~ 0 < (\mu/T)^2 <  \epsilon \}$
 and
 $C_2 = \{ \mu^2 \in \mathbb{R} ~|~ -\epsilon < (\mu/T)^2 < 0 \}$,
 where $\epsilon$ is a positive infinitesimal value.
If $F_1$ and $F_2$ have the connected domain at $\mu^2=0$,
 we can use the analytic continuation exactly.
In Scenarios A and B,
 there exists the singularity at $\mu^2=0$
 and hence the analytic continuation is no longer possible
 due to $F_1 \neq F_2$ for inhomogeneous chiral condensates.
In Scenario C, on the other hand, there is a possibility
 that the analytic continuation still works.

\subsubsection{Scenarios A and B}

In Scenario A, the $\mu^2=0$ line becomes singular,
 so that the analytic continuation is impossible.
This situation, however, becomes non-trivial
 when a Taylor expansion
 in terms of $\mu_\mathrm{R}$ at $\mu^2=0$
 is considered.
The Taylor series cannot reproduce the non-analyticity
 and hence can describe only one side of the solution.
To choice which solution is realized
 (whether the homogeneous or inhomogeneous
 solutions),
 we need a restriction to the Taylor series,
 since we cannot mathematically determine
 the Taylor series just on such a singular point.
In Scenario B, the situation is completely the same as Scenario A,
 where the $\mu^2=0$ line is the singular line
 due to $\mathrm{Im}~\!Z \neq 0$ at finite $\mu_\mathrm{I}$.
The Taylor series with the restriction, $\mathrm{Im}~\!Z = 0$,
 may lead to solutions continued to the $\mu_\mathrm{R}$ region.
However, such a series is not mathematically well-defined.
We note here that one can consider a first-order transition at
 $\mu^2=0$, where $q^2$ has a finite positive value in the $\mu^2
 \to -0$ limit,
 but our conclusion is not changed in this scenario.

\subsubsection{Scenario C}

In Scenario C, the wavenumber has a nonzero real value,
 $q_\mathrm{R} \neq 0$, in the $C_1$ region,
 while in the $C_2$ region it becomes pure imaginary,
 $q_{\rm I} \neq 0$.
Here $Z$ has only the q-even terms at $\mu^2 = 0$,
 since $\alpha_{2n+1}$ is the odd function of $\mu$
 in the expansion (\ref{gl}).
Now we can further consider two situations, (I) and (II):
\begin{description}
\item[Scenario C(I) ---]
	Assuming that $q$ vanishes in the $\mu^2 \rightarrow +0$ limit,
	$\alpha_2 (T,\mu)$ should be positive at $\mu = 0$.
	In the $\mu^2 \rightarrow -0$ limit, on the other hand,
	the coefficient of $q^2_{\rm I}$
	should be $-\alpha_2 < 0$ for $q=iq_{\rm I}$.
	Thus, the point, $q_{\rm I}=0$, cannot be a minimum of $Z$,
	and $q_{\rm I}$ does not vanish in the $\mu^2 \rightarrow -0$ limit.
	In this case, there is the first-order phase transition at $\mu^2=0$.
	However, there is no reason that the translational (rotational) symmetry
	is spontaneously broken in the $\mu^2 \to -0$ limit as well.
\item[Scenario C(II) ---]
	In Scenario C(I)
	it is assumed that $\alpha_2 (T,\mu=0) \neq 0$,
	but $\alpha_2=0$ can be acceptable.
	In this case, it seems possible to perform the analytic continuation
	because $q^2$ in the $\mu^2 \to \pm 0$ limit
	can be smooth at $\mu^2=0$.
      This situation is quite similar to the quark number density, $n_q$,
	in the system with $B=0$,
	since $n_q^2$ is positive (negative)
	at finite $\mu_\mathrm{R}$ ($\mu_\mathrm{I}$)
	and $n_q^2$ is smoothly connected at $\mu^2=0$
        when we take the $\mu^2 \to \pm 0$ limit.
\end{description}
In Scenario C(I), the analytic continuation is impossible
 due to the first-order transition at $\mu^2=0$.
Although the Taylor series constructed just on the singular point
 is not mathematically well-defined,
 each solution of inhomogeneous condensates may be reproduced
 under a suitable choice of input configurations.
However, this situation is not realistic,
 since there is no reason to consider that
 the spontaneous translational (rotational) symmetry
 breaking occurs at $\mu^2=0$.
Thus, we can exclude this situation.
In Scenario C(II), on the other hand,
 there has the possibility that the analytic continuation works
 because $q^2$ becomes smooth at $\mu^2=0$
 as same as $n_q^2$ in the system with vanishing $B$.
Only this scenario
 has the well-defined analytic continuation process
 and the Taylor series constructed at $\mu^2=0$.

If inhomogeneous chiral condensates appear on the QCD phase diagram,
 the singular line, $\mu^2=0$, should be exist.
Note that this singular line can appear at smaller $\mu_\mathrm{R}/T$
 than the natural boundary of the Taylor series in the absence of
 the magnetic field or the existence of the nonzero current mass.
To avoid such a singularity in the analytic continuation,
 we can consider the complex chemical potential.
However,
 lattice QCD simulations and effective model calculations
 are quite difficult, since the singular region should have nonzero
 $\mathrm{Im}~\!Z$.

\subsection{Non-analyticity and dense lattice QCD simulations}

Now we discuss the consequences of the non-analyticity
 for several methods to work around the sign problem
 used in the lattice QCD simulations at finite chemical potentials.

The trivial case is the analytic continuation method
 from $\mu_\mathrm{I}$ to $\mu_\mathrm{R}$%
~\cite{deForcrand:2002ci,deForcrand:2003hx,D'Elia:2002gd,D'Elia:2004at,Chen:2004tb}.
This is nothing but our discussion about the analytic continuation.
Thus, it is not feasible.

The reweighting method%
~\cite{Fodor:2001au,Fodor:2002km,Fodor:2001pe,Fodor:2004nz}
 can go beyond the non-analytic point in principle.
However, the overlap problem is important in this method.
The $\mu_\mathrm{I}$ region cannot be used
 for the creation of the probability in the important sampling procedure,
 since in Scenario A the overlap which can induce $q \neq 0$
 should be vanished or very small,
 and there is the sign problem at finite $\mu_\mathrm{I}$ in Scenario B.
Also, the $\mu^2=0$ point may have a small overlap problem.
If we know regions
 where the sign problem does not arise
 and inhomogeneous chiral condensates exist,
 the reweighting method should completely work,
 while we do not know such a convenient region at present.

The canonical approach%
~\cite{Hasenfratz:1991ax,Alexandru:2005ix,Kratochvila:2006jx,deForcrand:2006ec,Li:2010qf}
 is based on the fact that the grand canonical partition function
 with $\mu_\mathrm{R}$ can be constructed
 from the one with $\mu_\mathrm{I}$,
 which is valid if the quark number density is a good quantum number.
In this method,
 we do not directly use the analytic continuation, so that
 the non-analyticity does not affect this method in principle.
However, it is quite difficult to pick up
 information on inhomogeneous chiral condensates,
 since some inhomogeneous properties should be hidden
 when we calculate the grand canonical partition function
 or some observables via the canonical partition function.
This fact should be consistent with the results of a Lee-Yang zero analysis
 in QCD~\cite{Barbour:1991vs,Nakamura:2013ska,Nagata:2014fra},
 since we only see the distribution of zeros
 and hence it is difficult to clarify the existence of inhomogeneous
 phases even if we can find any phase transitions
 from the behavior of zeros.

On the other hand, the Taylor expansion method
~\cite{Allton:2005gk,Gavai:2008zr} is nontrivial.
Since this method is deeply related to the analytic continuation method,
 the Taylor series plays a crucial role.
The Taylor series in the Taylor expansion method constructed at $\mu^2=0$
 is the holomorphic function.
Thus, the $\mu_\mathrm{R}$ and $\mu_\mathrm{I}$ regions
 cannot be simultaneously reproduced
 because there is the non-analyticity at $\mu^2=0$.
Mathematically, such Taylor series are not well-defined,
 but one can describe the both-side of solutions
 by considering the restriction to make
 the Taylor series describing the desirable limit by hand in the
 numerical calculations.
This procedure may be affected by the numerical error.

From the above discussion,
 we find that since the $\mu^2=0$ point becomes the singular point,
 extreme care should be taken to perform
 the lattice QCD simulations
 with the Taylor expansion method, the
 reweighting method, the canonical approach,
 and the analytic continuation method
 in the presence of the magnetic field.
If the effect of the non-analyticity at $\mu^2=0$ are not correctly treated,
 information about the inhomogeneous chiral condensates should be missed
 in the Taylor expansion and analytic continuation methods.
Such a non-analyticity can be taken into account
 by imposing the restriction mentioned above.
Therefore, the information missing problem should be carefully considered
 in the dense lattice QCD simulations.

The information missing problem can be avoided
 in the complex Langevin method
~\cite{Parisi:1980ys,Parisi:1984cs}
 and the Lefschetz-thimble path integral method
~\cite{Witten:2010cx,Cristoforetti:2012su,Fujii:2013sra},
 since the dense lattice QCD simulation can be directly performed
 at finite $\mu_\mathrm{R}$.
However, both methods are far from perfection:
 there is no guarantee that the complex Langevin method leads to
 the correct answer when logarithmic terms exist in the action
 (see, e.g.,
 Refs~\cite{Fujimura:1993cq,Mollgaard:2013qra,Aarts:2014nxa}, and
 Ref.~\cite{Nishimura:2015pba} from the viewpoint of the
 singular drift term).
 The Lefschetz-thimble path integral method can provide
 the correct answer in principle,
 but the actual lattice calculation method is not completed,
 e.g., how to include the multi-thimble contributions.

The effective model approaches, on the other hand,
 are free from the information missing problem.
The lattice QCD simulations
 without the assumption of inhomogeneous chiral condensates
 can be possible by the suitable choice of spatial boundary conditions.
If the effective models are constrained by such lattice QCD
 data at finite $\mu_\mathrm{I}$,
 one can obtain the reliable model
 for the system without inhomogeneous phases.
This method is nothing but the imaginary chemical potential matching
 approach~\cite{Sakai:2008um,Kashiwa:2008bq}.
However, information about inhomogeneous chiral condensed phases
 can be introduced to the effective models
 by assuming the shape of the solution.
This is a standard procedure to include the inhomogeneous condensates
 in the effective model approach.
By unifying the substitution procedure
 and the imaginary chemical potential matching approach,
 one can investigate {\it full QCD}.

\section{Summary}

In this paper, we have investigated
 the properties of inhomogeneous chiral condensates
 at finite imaginary chemical potential
 in the presence of an external magnetic field.
From the reality of the partition function and the symmetry
 argument,
 we have considered two possible scenarios
 for the DCDW at finite $\mu_\mathrm{I}$.
In both scenarios,
 the $\mu^2=0$ point becomes the singular,
 which is induced by the inhomogeneous condensates.
This singularity disturbs the investigation of
 inhomogeneous chiral condensates,
 and hence information of such condensates sometimes is missed,
 i.e., the information missing problem.
We found that the information missing problem can appear
 in the lattice QCD simulation
 with the Taylor expansion and analytic continuation methods.
Problems to observe the inhomogeneous chiral condensates
 by using the reweighting method and the canonical approach are also discussed.
Moreover, we considered the exceptional case where $q^2$ becomes negative.

In the analytic continuation,
 holomorphic functions, e.g., the Taylor series,
 play a crucial role.
If the $\mu^2=0$ point becomes
 the connected domain
 where two holomorphic functions prepared
 in the $\mu_\mathrm{R}$ and $\mu_\mathrm{I}$ regions coincide,
 the present information missing problem does not arise.
However, there exists the singularity at $\mu^2=0$,
 which is induced from the different form of chiral condensates
 or the imaginary part of the partition function.
Thus, the analytic continuation from the $\mu^2 \le 0$ region
 is not possible, since the line $\mu^2=0$ forms the singular line.
This fact leads to that the Taylor expansion method
 require an extreme care if the Taylor series is constructed at
 $\mu^2=0$, the singular point.
To correctly construct the Taylor series
 which continues to the $\mu_\mathrm{R}$ region,
 some restrictions are required in the numerical code.
It should be noted that
 we can still consider Scenario C(II)
 which has the well-defined analytic continuation.
However, even if which scenarios are realized,
 we must check the behavior of $q^2$ at $\mu^2 < 0$
 before investigating the $\mu^2 > 0$ region
 to clarify the existence of non-analyticity.

For finite current quark masses,
 the singular line can be sifted at finite $\mu_\mathrm{R}$ region,
 which leads to the possibility
 that the singular line appears at the smaller
 $\mu_\mathrm{R}/T$ than the natural boundary of the Taylor
 series.
In this case, the information missing problem becomes more serious.
Consequently, the applicable range of the Taylor expansion method
 and that of the analytic continuation method are strongly restricted.
It should be noted that
 we cannot estimate the position of the singular line
 by only using the lattice QCD data,
 unlike the natural boundary
 which can be estimated from the convergence behavior of the series.
Meanwhile, a nonzero current quark mass
 gives rise to the configuration change
 of the DCDW condensate~\cite{Karasawa:2013zsa}.

Possible promising methods
 to overcome the information missing problem at large $\mu_\mathrm{R}$
 would include the unification of
 the imaginary chemical potential matching approach
 and the inputting of the solutions of inhomogeneous chiral condensates.
In this case, one can obtain reliable effective models at finite
 $\mu_\mathrm{R}$ without the inhomogeneous phases,
 but the information on the inhomogeneous condensations
 can be restored by inputting the solution.
In this method,
 the lattice QCD data without inhomogeneous chiral condensates
 at finite $\mu_\mathrm{I}$ is required
 but can be obtained if the spatial boundary conditions are suitably imposed.
Therefore, we do not need details of
 information about the inhomogeneous condensates at finite $\mu_\mathrm{I}$.
Also, it is interesting to combine this approach with a nonperturbative method,
 e.g., the functional renormalization group method or the Dyson-Schwinger formalism.
These topics will be considered in future studies.

\vspace{2mm}
{\it Acknowledgments:}
The authors thank Keitaro Nagata, Hideaki Iida, and Toshitaka Tatsumi for helpful comments.
K.K. and R.Y. are supported by Grants-in-Aid for Japan Society for the Promotion
 of Science (JSPS) fellows No.26-1717 and No.27-1814, respectively.
T.-G.L. is partially supported by Grant-in-Aid for Scientific Research on Innovative Areas
 thorough No. 24105008 provided by MEXT.

\bibliography{ref.bib}

\end{document}